\documentclass[acmsmall]{acmart}
\usepackage{macros}

\setcopyright{none}
\begin{document}

\title{Josephine}
\subtitle{Using JavaScript to safely manage the lifetimes of Rust data}

\author{Alan Jeffrey}
\orcid{0000-0001-6342-0318}
\affiliation{
  \position{Research Engineer}
  \institution{Mozilla Research}
}
\email{ajeffrey@mozilla.com}

\begin{abstract}
This paper is about the interface between languages
which use a garbage collector and those which use fancy
types for safe manual memory management.
Garbage collection is the traditional memory management
scheme for functional languages, whereas type systems
are now used for memory safety in imperative languages.
We use existing techniques for linear capabilities
to provide safe access to copyable references,
but the application to languages
with a tracing garbage collector,
and to data with explicit lifetimes is new.
This work is related to mixed linear/non-linear
programming, but the languages being mixed are Rust and JavaScript.
\end{abstract}



\renewcommand\footnotetextcopyrightpermission[1]{}
\maketitle
\thispagestyle{empty}

\section{Introduction}

This paper is about the interface between languages
which use a garbage collector and those which use fancy
types for safe manual memory management.

Garbage collection is the most common memory management technique for
functional programming languages, dating back to LISP~\cite{LISP}.
Having a garbage collector guarantees memory safety, but at the
cost of requiring a runtime system.

Imperative languages often require the programmer to perform
manual memory management, such as the \verb|malloc| and \verb|free|
functions provided by C~\cite{K+R}. The safety of a program
(in particular the absence of \emph{use-after-free} errors)
is considered the programmer's problem.
More recently, languages such as Cyclone~\cite{cyclone}
and Rust~\cite{rust} have used fancy type systems
such as substructural types~\cite{girard,Go4,walker}
and region analysis~\cite{regions} to guarantee memory
safety without garbage collection.

This paper discusses the Josephine API~\cite{josephine} for using the
garbage collector provided by the SpiderMonkey~\cite{spidermonkey}
JavaScript runtime to safely manage the lifetime of Rust~\cite{rust}
data. It uses techniques from $L^3$~\cite{l3} and its application
to regions~\cite{l3-with-regions}, but the application to languages
with a tracing garbage collector, and to languages with explicit
lifetimes is new.

\subsection{Rust}

Rust is a systems programming language which uses fancy types to
ensure memory safety even in the presence of mutable update, and
manual memory management. Rust has an affine type system, which
allows data to be discarded but does not allow data to be arbitrarily
copied. For example, the Rust program:
\begin{verbatim}
  let hello = String::from("hello");
  let moved = hello;
  println!("Oh look {} is hello", moved);
\end{verbatim}
is fine, but the program:
\begin{verbatim}
  let hello = String::from("hello");
  let copied = hello;
  println!("Oh look {} is {}", hello, copied);
\end{verbatim}
is not, since \verb|hello| and \verb|copied| are simultaneously live. Trying to compile
this program produces:
\begin{verbatim}
  use of moved value: `hello`
   --> src/main.rs:4:32
    |
  3 |   let copied = hello;
    |       ------ value moved here
  4 |   println!("Oh look {} is {}", hello, copied);
    |                                ^^^^^ value used here after move
\end{verbatim}
The use of affine types allows aliasing to be tracked. For example, a
classic problem with aliasing is appending a string to itself. In
Rust, an example of appending a string is:
\begin{verbatim}
  let mut hello = String::from("hello");
  let ref world = String::from("world");
  hello.push_str(world);
  println!("Oh look hello is {}", hello);
\end{verbatim}
The important operation is \verb|hello.push_str(world)|, which mutates the
string \verb|hello| (hence the \verb|mut| annotation on the declaration of \verb|hello|).
The appended string \verb|world| is passed by reference,
(hence the \verb|ref| annotation on the declaration of \verb|world|).

A problem with mutably appending strings is ensuring that the string
is not appended to itself, for example the documentation for
C \verb|strcat| states ``Source and destination may not
overlap,'' but C does not check aliasing and relies on the programmer
to ensure correctness. In contrast, attempting to append a string
to itself in Rust:
\begin{verbatim}
  let ref mut hello = String::from("hello");
  hello.push_str(hello);
\end{verbatim}
produces an error:
\begin{verbatim}
  cannot borrow `*hello` as immutable because it is also borrowed as mutable
   --> src/main.rs:3:18
    |
  3 |   hello.push_str(hello);
    |   -----          ^^^^^- mutable borrow ends here
    |   |              |
    |   |              immutable borrow occurs here
    |   mutable borrow occurs here
\end{verbatim}
In Rust, the crucial invariant maintained by affine types is:
\begin{quote}\em
  Any memory that can be reached simultaneously by two different paths
  is immutable.
\end{quote}
For example in \verb|hello.push(hello)| there are two occurrences of \verb|hello| that
are live simultaneously, the first of which is mutating the string, so this is outlawed.

In order to track statically which variables are live simultaneously, Rust uses a lifetime
system similar to that used by region-based memory~\cite{regions}. Each allocation of
memory has a lifetime $\alpha$, and lifetimes are ordered $\alpha\subseteq\beta$.
Each code block introduces a lifetime, and for data which does not escape from its scope,
the nesting of blocks determines the ordering of lifetimes.

For example in the program:
\begin{verbatim}
  let ref x = String::from("hi");
  {
    let ref y = x;
    println!("y is {}", y);
  }
  println!("x is {}", x);
\end{verbatim}
the variable \verb|x| has a lifetime $\alpha$ given by the outer block,
and the variable \verb|y| has a lifetime $\beta\subseteq\alpha$ given by the inner block.

These lifetimes are mentioned in the types of references: the type $\REF\alpha T$
is a reference giving immutable access to data of type $T$, which will live at least as long as
$\alpha$. Similarly, the type $\REFMUT\alpha T$ gives mutable access to the data: the crucial
difference is that $\REF\alpha T$ is a copyable type, but $\REFMUT\alpha T$ is not.
For example
the type of \verb|x| is $\REF\alpha\STRING$ and the type of \verb|y| is
$\REF\beta(\REF\alpha\STRING)$, which is well-formed because $\beta\subseteq\alpha$.

Lifetimes are used to prevent another class of memory safety issues: use-after-free.
For example, consider the program:
\begin{verbatim}
  let hi = String::from("hi");
  let ref mut handle = &hi;
  {
    let lo = String::from("lo");
    *handle = &lo;
  } // lo is deallocated here
  println!("handle is {}", **handle);
\end{verbatim}
If this program were to execute, its behaviour would be undefined,
since \verb|**handle| is used after \verb|lo|
(which \verb|handle| points to) is deallocated. Fortunately, this program
does not pass Rust's borrow-checker:
\begin{verbatim}
  `lo` does not live long enough
   --> src/main.rs:6:11
    |
  6 |     *handle = &lo;
    |                ^^ borrowed value does not live long enough
  7 |   } // lo is deallocated here
    |   - `lo` dropped here while still borrowed
  8 |   println!("handle is {}", **handle);
  9 | }
    | - borrowed value needs to live until here
\end{verbatim}
This use-after-free error can be detected because (naming the outer lifetime as
$\alpha$ and the inner lifetime as $\beta\subseteq\alpha$) \verb|handle| has type
$\REFMUT\alpha\REF\alpha\STRING$, but \verb|&lo| only has type $\REF\beta\STRING$, no
$\REF\alpha\STRING$ as required by the assignment.

Lifetimes avoid use-after-free by maintaining two invariants:
\begin{quote}\em
  Any dereference happens during the lifetime of the reference, \\
  and deallocation happens after the lifetime of all references.
\end{quote}
There is more to the Rust type system than described here
(higher-order functions, traits, variance, concurrency, \dots) but the important features
are \emph{affine types} and \emph{lifetimes} for ensuring memory safety,
even in the presence of manual memory management.

\subsection{SpiderMonkey}

SpiderMonkey~\cite{spidermonkey} is Mozilla's JavaScript runtime, used in the Firefox browser,
and the Servo~\cite{servo} next-generation web engine. This is a full-featured
JS implementation, but the focus of this paper is its automatic memory management.

Inside a web engine, there are often native implementations of HTML features,
which are exposed to JavaScript using DOM interfaces. For example, an HTML image
is exposed to JavaScript as a DOM object representing an \verb|<img>| element,
but behind the scenes there is native code responsible for loading and rendering
images.

When a JavaScript object is garbage collected, a destructor is called to
deallocate any attached native memory. In the case that the native code
is implemented in Rust, this leads to a situation where Rust relies on affine
types and lifetimes for memory safety, but JavaScript respects neither of these.
As a result, the raw SpiderMonkey interface to Rust is very unsafe,
for example there are nearly 400 instances of unsafe code in the Servo
DOM implementation:
\begin{verbatim}
  $ grep "unsafe_code" components/script/dom/*.rs | wc
      393     734   25514
\end{verbatim}
Since JavaScript does not respect Rust's affine types,
Servo's DOM implementation makes use of Rust~\cite[\S3.11]{rust}
\emph{interior mutability} which replaces the compile-time type
checks with run-time dynamic checks. This carries run-time overhead,
and the possibility of checks failing, and Servo panicking.

Moreover, SpiderMonkey has its own invariants, and if an embedding
application does not respect these invariants, then runtime errors can
occur. One of these invariants is the division of JavaScript memory
into \emph{compartments}~\cite{compartments}, which can be garbage collected
separately. The runtime has a notion of ``current compartment'',
and the embedding application is asked to maintain two invariants:
\begin{itemize}
  \item whenever an object is used, the object is in the current compartment, and
  \item there are no references between objects which cross compartments.
\end{itemize}
In order for native code to interact well with the SpiderMonkey garbage collector,
it has to provide two functions:
\begin{itemize}
\item a \emph{trace} function, that given an object, iterates over all of the
  JavaScript objects which are reachable from it, and
\item a \emph{roots} function, which iterates over all of the JavaScript
  objects that are live on the call stack.
\end{itemize}
From these two functions, the garbage collector can find all of the reachable
JavaScript objects, including those reachable from JavaScript directly, and
those reached via native code. The Josephine interface to tracing
is discussed in \S\ref{sec:tracing}, and the interface to rooting
is discussed in \S\ref{sec:rooting}.

Automatically generating the trace function is reasonably straightforward
metaprogramming, but rooting safely turns out to be quite tricky.
Servo provides an approximate analysis for safe rooting using an ad-hoc
static analysis (the \emph{rooting lint}), but this is problematic because
a) the lint is syntax-driven, so does not understand about Rust features
such as generics, and b) even if it could be made sound it is disabled
more than 200 times:
\begin{verbatim}
  $ grep "unrooted_must_root" components/script/dom/*.rs | wc
      213     456   15961
\end{verbatim}

\subsection{Josephine}

Josephine~\cite{josephine} is intended to act as a safe bridge between
SpiderMonkey and Rust. Its goals are:
\begin{itemize}

\item to use JavaScript to manage the lifetime of Rust data,
  and to allow JavaScript to garbage collect unreachable data,

\item to allow references to JavaScript-managed data to be freely copied and discarded,
  relying on SpiderMonkey's garbage collector for safety,

\item to maintain Rust memory safety via affine types and lifetimes,
  without requiring additional static analysis such as the rooting lint,

\item to allow mutable and immutable access to Rust data via JavaScript-managed references,
  so avoiding interior mutability, and

\item to provide a rooting API to ensure that JavaScript-managed data is not garbage collected
  while it is being used.

\end{itemize}
Josephine is intended to be safe, in that any programs built using Josephine's API
do not introduce undefined behaviour or runtime errors.
Josephine achieves this by providing controlled access to
SpiderMonkey's \emph{JavaScript context}, and maintaining invariants about it:
\begin{itemize}

\item immutable access to JavaScript-managed data requires immutable access
  to the JavaScript context,

\item mutable access to JavaScript-managed data requires mutable access
  to the JavaScript context, and

\item any action that can trigger garbage collection (for example allocating
  new objects) requires mutable access to the JavaScript context.

\end{itemize}
As a result, since access to the JavaScript context does respect
Rust's affine types, mutation or garbage collection cannot occur
simultaneously with accessing JavaScript-managed data.

In other words, Josephine treats the JavaScript context as an affine
access token, or capability, which controls access to the JavaScript-managed
data. The data accesses respect affine types, even though the JavaScript objects
themselves do not.

This use of an access token to safely access data in a substructural
type system is \emph{not} new, it is the heart of Ahmed, Fluet and
Morrisett's $L^3$ Linear Language with Locations~\cite{l3} and its
application to linear regions~\cite{regions}.
Moreover, type systems for mixed linear/non-linear programming have
been known for more than 20 years~\cite{mixed}.

Other integrations of GC with linear types include Linear Lisp~\cite{linear-lisp},
Alms~\cite{alms},
Linear Haskell~\cite{linear-haskell}, and
linear OCaml~\cite{linear-ocaml}, but these do not integrate with Rust's
lifetime model.

Garbage collection for Rust has previously been investigated,
e.g.~in Servo~\cite{servo-gc} or the rust-gc library~\cite{rust-gc},
but these approaches take a different approach: in Servo, the
API by itself is unsafe and depends on interior mutability and
a separate rooting lint for safety. The rust-gc library uses
reference counting and interior mutability. Neither of them interact
with lifetimes in the way Josephine does.

The aspects of Josephine that are novel are:
\begin{itemize}

\item the languages being mixed are Rust and 
  JavaScript, which are both industrial-strength,

\item the treatment of garbage collection requires
  different typing rules than regions in $L^3$, and

\item the types for JS-managed references respect the Rust
  lifetime system.

\end{itemize}

\subsection*{Acknowledgments}

This work benefited greatly from conversations with
Amal Ahmed,
Nick Benton,
Josh Bowman-Matthews,
Manish Goregaokar,
Bobby Holly, and
Anthony Ramine.

\section{The Josephine API}

There are two important concepts in Josephine's API: \emph{JS-managed} data,
and the JS \emph{context}. For readers familiar with the region-based
variant~\cite{l3-with-regions} of $L^3$~\cite{l3}, JS-managed data
corresponds to $L^3$ references, and JS contexts to $L^3$ capabilities.

\subsection{JS-managed data}

JS-managed data has the type $\JSManaged{\alpha, C, T}$, which represents
a reference to data whose lifetime is managed by JS, which:
\begin{itemize}

\item is guaranteed to live at least as long as $\alpha$,
\item is allocated in JS compartment $C$, and
\item points to native data of type $T$.
  
\end{itemize}
This type is copyable, so not subject to the affine type discipline,
even though it can be used to gain mutable access to the native
data. We shall see later that this is safe for the same reason as
$L^3$: we are using the JS context as a capability, and it is not
copyable.

In examples, we make use of Rust's \emph{lifetime elision}~\cite[\S3.4]{rustinomicon},
and just write $\JSManaged{C,T}$ where the lifetime $\alpha$ can be
inferred.

In the simplest case, $T$ is a base type like $\STRING$, but in more complex
cases, $T$ might itself contain JS-managed data, for example a type of
cells in a doubly-linked list can be defined:
\begin{verbatim}
  type Cell<'a, C> = JSManaged<'a, C, NativeCell<'a, C>>;
\end{verbatim}
where:
\begin{verbatim}
  struct NativeCell<'a, C> {
    data: String,
    prev: Option<Cell<'a, C>>,
    next: Option<Cell<'a, C>>,
  }
\end{verbatim}
This pattern is a common idiom, in that there are two types:
\begin{itemize}
\item $\NativeCell{\alpha,C}$ containing the native representation
of a cell, including the prev and next
references, and
\item $\Cell{\alpha,C}$ containing a reference to a native cell,
whose lifetime is managed by JS.
\end{itemize}
These types are both parameterized by a lower bound $\alpha$ on the lifetime
of the cell, and the compartment $C$ that the cell lives in.

Doubly-linked lists are an interesting example of programming in Rust,
and indeed there is an introductory text \emph{Learning Rust With
  Entirely Too Many Linked Lists}~\cite{too-many-lists}, in which safe
implementations of doubly-linked lists require interior mutability
(and hence dynamic checks) and reference counting.

\subsection{The JS context}

By itself, JS-managed references are not much use: there has to be an
API for creating and dereferencing them: this is the role of the
JS \emph{context}, which acts as a capability for manipulating
JS-managed data. The JS context is part of the SpiderMonkey API,
where it is used to store state that is global to the runtime system.

There is only one JS context per thread (and JS contexts cannot be shared
or sent between threads) so unique access to the JS context implies unique
access to all JS-managed data. We can use this to give safe mutable access
to JS-managed data, since the JS context is a unique capability.

The JS context has a state, notably keeping track of the current
compartment, but also permissions such as ``allowed to create new
references'' or ``allowed to dereference''.  This state is tracked
using phantom types, so the JS context
has type $\JSContext{S}$, where $S$ is the current state.

For example, a program to allocate a new JS-managed reference is:
\begin{verbatim}
  let x: JSManaged<C, String> = cx.manage(String::from("hello"));
\end{verbatim}
and a program to access a JS-managed reference is:
\begin{verbatim}
  let msg: &String = x.borrow(cx);
\end{verbatim}
These programs make use of the JS context \verb|cx|. In order for the
first example to typecheck:
\begin{itemize}

\item \verb|cx| must have type $\REFMUT{}\JSContext{S}$, where
\item $S$ (the state of the context) must have permission to allocate
  references in $C$, and
\item $C$ must be a compartment.

\end{itemize}
The second example is similar, except:
\begin{itemize}

\item we do not need mutable access to the context, and
\item $S$ must have permission to access compartment $C$.

\end{itemize}
Fortunately, Rust has a \emph{trait} system (similar to Haskell's
class system), which allows us to express these constraints.  In the
same way that $C$ and $S$ are phantom types, these are \emph{marker}
traits with no computational value. The typing for
the first example is:
\begin{verbatim}
  (cx: &mut JSContext<S>) where
    S: CanAlloc + InCompartment<C>,
    C: Compartment,
\end{verbatim}
and for the second:
\begin{verbatim}
  (cx: &JSContext<S>) where
    S: CanAccess,
    C: Compartment,
\end{verbatim}
A program to mutably access a JS-managed reference is:
\begin{verbatim}
  let msg: &mut String = x.borrow_mut(cx);
\end{verbatim}
at which point the fact that the JS context is an affine capability
becomes important. The typing required for this is:
\begin{verbatim}
  (cx: &mut JSContext<S>) where
    S: CanAccess,
    C: Compartment,
\end{verbatim}
That is \emph{unique access to JS-managed data requires unique access to the JS context},
and so we cannot simultaneously have mutable access to two different JS-managed
references. This is the same safety condition that region-based $L^3$ uses.

For example, we can use this (together with Rust's built-in replace function
which swaps the contents of a mutable reference) to replace the contents of a cell
with a new value:
\begin{verbatim}
  fn replace<S>(self, cx: &'a mut JSContext<S>, new_data: String) -> String where
    S: CanAccess,
    C: Compartment,
  {
    let ref mut old_data = self.0.borrow_mut(cx).data;
    replace(old_data, new_data)
  }
\end{verbatim}

\subsection{Typing access}
\label{sec:typing-access}

A first-cut type rule for accessing data in a typing context
in which $S: \CanAccess$ and $C: \Compartment$ is:
\begin{quote}
  if~$cx: \REF{} \JSContext{S}$ and $p: \JSManaged{C,T}$
  then $p.\borrow(cx): \REF{} T$
\end{quote}
(and similarly for mutable access)
which is fine, but does not mention the lifetimes. Adding these in gives
the type rule:
\begin{quote}
  if~$cx: \REF{\alpha} \JSContext{S}$ and $p: \JSManaged{\alpha,C,T}$
  then $p.\borrow(cx): \REF\alpha T$
\end{quote}
which is correct, but assumes that the lifetime that the JS context
has been borrowed for is exactly the same as the lifetime of the
reference. Separating these gives (when $\alpha \subseteq \beta$):
\begin{quote}
  if~$cx: \REF{\alpha} \JSContext{S}$ and $p: \JSManaged{\beta,C,T}$
  then $p.\borrow(cx): \REF\alpha T$
\end{quote}
This rule is still incorrect, but for a slightly subtle reason.
It \emph{is} correct when $T$ is a base type, but fails in the case of
a type which includes nested JS-managed references. If that were the
rule, then we could write programs such as
\begin{verbatim}
   let cell: Cell<'a, C> = ...;
   let next: Cell<'a, C> = cell.borrow(cx).next?; // cell is keeping next alive
   cell.borrow_mut(cx).next = None;            // nothing is keeping next alive
   cx.gc();                                       // something that triggers GC
   next.borrow(cx);                                 // this is a use-after-free
\end{verbatim}
The problem in this example is that after setting cell's next pointer to \verb|None|,
there is nothing in JS keeping \verb|next| alive, so it is reachable
from Rust but not from JS. After a GC, the JS runtime can deallocate
\verb|next|, so accessing it is a use-after-free error.

In a language with built-in support for GC, there would be a hidden
GC root introduced by putting \verb|next| on the stack, but Rust does
not have support for such hidden rooting.

The problem in general is that when accessing $p:\JSManaged{\beta,C,T}$,
using a JS context borrowed
for lifetime $\alpha\subseteq\beta$, there may be nested JS-managed
data, also with lifetime $\beta$. These are being kept alive by $p$,
which is fine as long as $p$ is not mutated, but mutating $p$
might cause them to become unreachable in JS, and thus candidates
for garbage collection.

The fix used by Josephine is to replace any nested uses of
$\beta$ in $T$ by $\alpha$, that is the type rule is
(when $\alpha \subseteq \beta$):
\begin{quote}
  if~$cx: \REF{\alpha} \JSContext{S}$ and $p: \JSManaged{\beta,C,T}$
  then $p.\borrow(cx): \REF\alpha T[\alpha/\beta]$
\end{quote}
The conjecture that Josephine makes is that this is safe, because
GC cannot happen during the lifetime $\alpha$. In order to ensure
this, we maintain an invariant:
\begin{quote}\em
  Any operation that can trigger garbage collection
  requires mutable access to the JS context.
\end{quote}
This is why \verb|cx.manage(data)| requires
\verb|cx| to have type $\REFMUT{}\JSContext{S}$, \emph{not} because
we are mutating the JS context itself, but because allocating
a new reference might trigger GC.

In Rust, the substitution $T[\alpha/\beta]$ is expressed by
$T$ implementing a trait $\JSLifetime{\alpha}$:
\begin{verbatim}
  pub unsafe trait JSLifetime<'a> {
    type Aged;
    unsafe fn change_lifetime(self) -> Self::Aged { ... }
  }
\end{verbatim}
This is using an \emph{associated type} $T\cc\Aged$
to represent $T[\alpha/\beta]$.
In particular, $\JSManaged{\beta,C,T}$ implements
$\JSLifetime{\alpha}$ as long as $T$ does,
and $\JSManaged{\beta,C,T}[\alpha/\beta]$ is $\JSManaged{\alpha,C,T[\alpha/\beta]}$.

The implementation of $\JSLifetime{\alpha}$ has a lot of boilerplate,
but fortunately that boilerplate is amenable to Rust's metaprogramming
system, so user-defined types can just mark their type as
\verb|#[derive(JSLifetime)]|.

\section{Interfacing to the garbage collector}

Interfacing to the SpiderMonkey GC has two parts:
\begin{itemize}

\item \emph{tracing}: from a JS-managed reference, find
  the JS-managed references immediately reachable from it, and

\item \emph{rooting}: find the JS-managed references which
  are reachable from the stack.

\end{itemize}
From these two functions, it is possible to find all of the
JS-managed references which are reachable from Rust. Together
with SpiderMonkey's regular GC, this allows the runtime to
find all of the reachable JS objects, and then to reclaim the
unreachable ones.

These interfaces are important for program correctness, since
under-approximation can result in use-after-free,
and over-approximation can result in space leaks.

In this section, we discuss how Josephine supports these interfaces.

\subsection{Tracing}
\label{sec:tracing}

Interfacing to the SpiderMonkey tracer via Josephine is achieved
in the same way as Servo~\cite{servo},
by implementing a trait:
\begin{verbatim}
  pub unsafe trait JSTraceable {
    unsafe fn trace(&self, trc: *mut JSTracer);
  }
\end{verbatim}
Josephine provides an implementation:
\begin{verbatim}
  unsafe impl<'a, C, T> JSTraceable for JSManaged<'a, C, T> where
    T: JSTraceable { ... }
\end{verbatim}
User-defined types can then implement the interface by
recursively visiting fields, for example:
\begin{verbatim}
  unsafe impl<'a, C, T> JSTraceable for NativeCell<'a, C> {
    unsafe fn trace(&self, trc: *mut JSTracer) {
      self.prev.trace(trc);
      self.next.trace(trc);
    }
  }
\end{verbatim}
This is a lot of unsafe boilerplate, but fortunately can
also be mechanized using meta-programming by marking a type
as \verb|#[derive(JSTraceable)]|.

One subtlety is that during tracing data of type $T$, the JS runtime
has a reference of type $\REF{}{T}$ given by the \verb|self| parameter
to \verb|trace|. For this to be safe, we have to ensure that there is
no mutable reference to that data. This is maintained by the
previously mentioned invariant:
\begin{quote}\em
  Any operation that can trigger garbage collection
  requires mutable access to the JS context.
\end{quote}
Tracing is triggered by garbage collection, and so had unique access
to the JS context, so there cannot be any other live mutable access
to any JS-managed data.

\subsection{Rooting}
\label{sec:rooting}

In languages with native support for GC, rooting is
supported by the compiler, which can provide metadata for
each stack frame allowing it to be traced. In languages like
Rust that do not have a native GC, this metadata is not
present, and instead rooting has to be performed explicitly.

This explicit rooting is needed whenever an object is
needed to outlive the borrow of the JS context that produced
it. For example, a function to insert a new cell after
an existing one is:
\begin{verbatim}
  pub fn insert<C, S>(cell: Cell<C>, data: String, cx: &mut JSContext<S>) where
    S: CanAccess + CanAlloc + InCompartment<C>,
    C: Compartment,
  {
    let old_next = cell.borrow(cx).next;
    let new_next = cx.manage(NativeCell {
      data: data,
      prev: Some(cell),
      next: old_next,
    });
    cell.borrow_mut(cx).next = Some(new_next);
    if let Some(old_next) = old_next {
      old_next.borrow_mut(cx).prev = Some(new_next);
    }
  }
\end{verbatim}
This is the ``code you would first think of'' for inserting an
element into a doubly-linked list, but is in fact not safe because
the local variables \verb|old_next| and \verb|new_next| have not been
rooted. If GC were triggered just after \verb|new_next| was created,
then it could be reclaimed, causing a later use-after-free.

Fortunately, Josephine will catch these safety problems, and report
errors such as:
\begin{verbatim}
  error[E0502]: cannot borrow `*cx` as mutable because
    it is also borrowed as immutable
    |
    |         let old_next = self.borrow(cx).next;
    |                                    -- immutable borrow occurs here
    |         let new_next = cx.manage(NativeCell {
    |                        ^^ mutable borrow occurs here
...
    |     }
    |     - immutable borrow ends here
\end{verbatim}
The fix is to explicitly root the local variables. In Josephine this is:
\begin{verbatim}
   let ref mut root1 = cx.new_root();
   let ref mut root2 = cx.new_root();
   let old_next = (... as before ...).in_root(root1);
   let new_next = (... as before ...).in_root(root2);
\end{verbatim}
The declaration of a \verb|root| allocates space on the stack
for a new root, and \verb|managed.in_root(root)| roots \verb|managed|.
Note that it is just the reference that is copied to the stack,
the JS-managed data itself doesn't move.
Roots have type $\JSRoot{\beta,T}$ where $\beta$ is the lifetime
of the root, and $T$ is the type being rooted.

Once the local variables are rooted, the code typecheck,
because rooting changes the lifetime of the JS-managed
data, for example (when $\alpha \subseteq \beta$):
\begin{quote}
  if~$p: \JSManaged{\alpha,C,T}$ \\
  and~$r: \REFMUT{\beta}{\JSRoot{\beta,\JSManaged{\beta,C,T[\beta/\alpha]}}}$ \\
  then~$p.\inRoot(r): \JSManaged{\beta,C,T[\beta/\alpha]}$.
\end{quote}
Before rooting, the JS-managed data had lifetime $\alpha$,
which is usually the lifetime of the borrow of the JS context
that created or accessed it.
After rooting, the JS-managed data has lifetime $\beta$,
which is the lifetime of the root itself. Since roots are
considered reachable by GC, the contents of a root
are guaranteed not to be GC'd during its lifetime,
so this rule is sound.

Note that this use of substitution $T[\beta/\alpha]$
is being used to extend the lifetime of the
JS-managed data, since $\alpha\subseteq\beta$. This
is in comparison to the use of substitution $T[\alpha/\beta]$
in \S\ref{sec:typing-access}, which was used to contract the
lifetime.

\section{Compartments}

SpiderMonkey uses \emph{compartments} to organize memory,
so that garbage collection does not have to sweep the
entire memory, just one compartment\footnote{%
  For purposes of this paper, we are ignoring the distinction between zones and
  compartments.
}
To achieve this, SpiderMonkey maintains the invariant:
\begin{quote}\em
  There are no direct references between compartments.
\end{quote}
This invariant is expected to be maintained by
any native data: tracing a JS-managed object
should never result in tracing an object from a different
compartment.

In Josephine, the compartment that
native data has been placed in is part of its
type. Data of type $\JSManaged{C,T}$ is attached
to a JS object in compartment $C$.

\subsection{Maintaining the invariant}

It would be possible for user-defined types to break
the compartment invariant, for example:
\begin{verbatim}
  type BadCell<'a, C, D> = JSManaged<'a, C, NativeBadCell<'a, C, D>>;
\end{verbatim}
where:
\begin{verbatim}
  struct NativeBadCell<'a, C, D> {
    data: String,
    prev: Option<BadCell<'a, C, C>>,
    next: Option<BadCell<'a, D, D>>,
  }
\end{verbatim}
This type violates the compartment invariant, because
a cell of type \verb|BadCell<'a, C, D>| is in compartment
\verb|C| but its next pointer is in compartment \verb|D|.

To maintain the compartment invariant, we introduce
a trait similar to \verb|JSLifetime|, but for compartments:
\begin{verbatim}
  pub unsafe trait JSCompartmental<C, D> {
    type ChangeCompartment;
  }
\end{verbatim}
In the same way that $\JSLifetime{\alpha}$ is used to implement
lifetime substitution $T[\alpha/\beta]$, the trait $\JSCompartmental{C,D}$
is used to implement compartment substitution $T[D/C]$. A type $T$ implementing
$\JSCompartmental{C,D}$ is asked to ensure that:
\begin{itemize}

\item $T$ is in compartment $C$,
\item $T$ only contains references to other types implementing $\JSCompartmental{C,D}$, and
\item $T\cc\ChangeCompartment$ is $T[D/C]$.

\end{itemize}
If the implementation of this type is incorrect, there may be safety issues,
which is why the trait is marked as \verb|unsafe|. Fortunately, deriving an
implementation of this trait is straightforward meta-programming.
Josephine provides a \verb|#[derive(JSCompartmental)|
which is guaranteed to maintain
the compartment invariant.

\subsection{Creating compartments}

In SpiderMonkey, a new compartment is created each time a
\emph{global object}~\cite[\S18]{ecmascript} is created.

In Josephine, there are two functions: one to create a new compartment,
and another to attach native data to the global. The global object can
be accessed with \verb|cx.global()|. For example:
\begin{verbatim}
  let cx = cx.create_compartment();
  let name = String::from("Alice");
  let cx = cx.global_manage(name);
\end{verbatim}
In some cases, the global (in freshly created compartment $C$)
contains some JS-managed data (also in compartment $C$), which is why
the initialization is split into two steps. First, create compartment
compartment $C$, then initialize the native data, which may make
use of $C$. For example:
\begin{verbatim}
   let cx = cx.create_compartment();
   let ref mut root = cx.new_root();
   let name = cx.manage(String::from("Alice")).in_root(root);
   let cx = cx.global_manage(NativeMyGlobal { name: name });
\end{verbatim}
where:
\begin{verbatim}
  struct NativeMyGlobal<'a, C> { name: JSManaged<'a, C, String> }
  type MyGlobal<'a, C> = JSManaged<'a, C, NativeMyGlobal<'a, C>>;
\end{verbatim}
The type rule for creating a compartment is:
\begin{quote}
  if $cx:\REFMUT{\alpha}{\JSContext{S}}$
  and $S: \CanAlloc + \CanAccess$ \\
  then $cx.\createCompartment() : \JSContext{S'}$ \\\mbox{}\quad
  where $S': \alpha + \CanAlloc + \InCompartment{C} + \IsInitializing{\alpha,C,T}$ \\\mbox{}\qquad
  for fresh $C: \Compartment$.
\end{quote}
Note:
\begin{itemize}

\item this is the first type rule which has changed the state of
  the JS context from $S$ to $S'$,

\item although $S$ can access data, $S'$ cannot: this is necessary for
  safety, since the global does not yet have any data attached to it,
  so accessing it would be undefined behaviour,

\item $S'$ only has lifetime $\alpha$, so we do not have two JS contexts
  live simultaneously,

\item $S'$ has entered the compartment $C$, and has the permission to create
  new objects in $C$,
  
\item $S'$ has the permission $\IsInitializing{C,T}$, which allows the global
  to be initialized with native data of type $T$.
  
\end{itemize}
The type rule for initializing a compartment is:
\begin{quote}
  if $cx: \JSContext{S}$
  and $x: T$
  and $S: \IsInitializing{\alpha,C,T}$ \\
  then $cx.\globalManage(x) : \JSContext{S'}$ \\\mbox{}\quad
  where $S': \alpha + \CanAccess + \CanAlloc + \InCompartment{C}$
\end{quote}

\subsection{Entering a compartment}
\label{sec:compartment-entering}

Given a JS-managed reference \verb|x|, we can enter its compartment with \verb|cx.enter_known_compartment(x)|.
This returns a JS context whose current compartment is that of \verb|x|.
For example, given a JS-managed reference \verb|x|,
we can create a new JS-managed reference in the same compartment with:
\begin{verbatim}
  let ref mut cx = cx.enter_known_compartment(x);
  let y = cx.manage(String::from("hello"));
\end{verbatim}
This has type rule:
\begin{quote}
  if $cx: \REFMUT{\alpha}{\JSContext{S}}$
  and $x: \JSManaged{C,T}$ \\
  and $S: \CanAccess + \CanAlloc$
  and $C: \Compartment$ \\
  then $cx.\enterKnownCompartment(x) : \JSContext{S'}$ \\\mbox{}\quad
  where $S': \alpha + \CanAccess + \CanAlloc + \InCompartment{C}$
\end{quote}

\subsection{Wildcard compartments}
\label{sec:compartment-wildcard}

Working with named compartments is fine when there is a fixed number of
them, but not when the number of compartments is unbounded. For
example, the type \verb|Vec<JSManaged<C, T>>| contains a vector of managed
objects, all in the same compartment, but sometimes you need a vector
of objects in different compartments. This is the same problem that
existential polymorphism~\cite{expoly}, and in particular
Java wildcards~\cite[\S8.1.2]{jls} is designed to solve, and we adopt
a similar approach.

The wildcard is called \verb|SOMEWHERE|, which we will
often abbreviate as $\SOMEWHERE$.
$\JSManaged{\SOMEWHERE, T}$ refers to JS-managed data whose
compartment is unknown. For example \verb|Vec<JSManaged<?, T>>|
contains a vector of managed objects, which may all be in different
compartments.

To create a wildcard, we use $x.\forgetCompartment()$, with type
rule:
\begin{quote}
  if $x: \JSManaged{C,T}$ 
  then $x.\forgetCompartment(): \JSManaged{\SOMEWHERE,T[\SOMEWHERE/C]}$
\end{quote}
Entering a wildcard compartment is the same as for a known compartment,
but renames the compartment to a fresh name:
\begin{quote}
  if $cx: \REFMUT{\alpha}{\JSContext{S}}$
  and $x: \JSManaged{\SOMEWHERE,T}$ \\
  and $S: \CanAccess + \CanAlloc$ \\
  then $cx.\enterUnknownCompartment(x) : \JSContext{S'}$ \\\mbox{}\quad
  where $S': \alpha + \CanAccess + \CanAlloc + \InCompartment{D}$ \\\mbox{}\qquad
  for fresh $D: \Compartment$.
\end{quote}
We also have a function $cx.\entered()$ of type $\JSManaged{D,T[D/\SOMEWHERE]}$,
which gives access to $x$ in its newly named compartment.
Note that access to data in a wildcard compartment is not allowed
(in the type system this is enforced since we do not have $\SOMEWHERE:\Compartment$),
for example:
\begin{verbatim}
  fn example<S>(cx: &mut JSContext<S>, x: JSManaged<SOMEWHERE, String>) where
    S: CanAccess,
  {
    // We can't access x without first entering its compartment.
    // Commenting out the next two lines gives an error
    // the trait `Compartment` is not implemented for `SOMEWHERE`.
    let ref mut cx = cx.enter_unknown_compartment(x);
    let x = cx.entered();
    println!("Hello, {}.", x.borrow(cx));
  }
\end{verbatim}

\section{Conclusions}

The contributions of this work are:
\begin{itemize}

\item an implementation of the ideas in $L^3$~\cite{l3} to mixed
  linear/non-linear programming~\cite{mixed}, where the
  languages being mixed are Rust and JavaScript,

\item a treatment of garbage collection (rather than region-based
  memory management) for such a system, and

\item a treatment of how operations such as accessing, mutating, and
  rooting can change the lifetimes of objects.
  
\end{itemize}
The main item left for future work is formalizing the approach
described here: memory safety is conjectured, but not proved
formally.

There are some aspects of the API which need more investigation:
\begin{itemize}

\item Other JavaScript engines take a different approach to
  rooting, notably V8 \emph{handle scopes}~\cite{v8-embedding},
  which have different trade-offs. In terms of this paper, the
  roots are attached to the JS context, rather than stored
  on the stack. It would be interesting to compare these approaches.

\item Josephine uses phantom types to track which compartment
  memory is allocated in, but does not support features such
  as \emph{cross-compartment wrappers}~\cite{compartments},
  which allow references between compartments.

\item In this paper, we have just used the SpiderMonkey runtime
  engine for its garbage collector, but it is a full-featured
  JavaScript engine, and it would be good to provide safe
  access to executing JS code. This would be simpler to achieve
  if there were a JS type system to generate bindings from,
  such as TypeScript~\cite{typescript}.
  
\end{itemize}
The distribution includes some simple examples such as doubly-linked lists
and a stripped-down DOM, but more examples are needed to see if the API
is usable for practical code.

\bibliographystyle{plain}
\bibliography{paper.bib}

\begin{thebibliography}{10}

\bibitem{l3}
A.~Ahmed, M.~Fluet, and G.~Morrisett.
\newblock L3: A linear language with locations.
\newblock {\em Fundamenta Informaticae}, 77(4):397--449, 2007.

\bibitem{linear-lisp}
H.~G. Baker.
\newblock Lively linear {Lisp}: ``look ma, no garbage!''.
\newblock {\em {SIGPLAN} Not.}, 27(8):89--98, 1992.

\bibitem{too-many-lists}
A.~Beingessner et~al.
\newblock {\em Learning {Rust} With Entirely Too Many Linked Lists}.
\newblock \url{http://cglab.ca/~abeinges/blah/too-many-lists/book/}.

\bibitem{mixed}
N.~Benton.
\newblock A mixed linear and non-linear logic: Proofs, terms and models.
\newblock In {\em Proc. Computer Science Logic}, pages 121--135, 1995.

\bibitem{Go4}
P.~N. Benton, G.~M. Bierman, V.~de~Paiva, and M.~Hyland.
\newblock A term calculus for intuitionistic linear logic.
\newblock In {\em Proc. Int. Conf. Typed Lambda Calculi and Applications},
  pages 75--90, 1993.

\bibitem{linear-haskell}
J.-P. Bernardy, M.~Boespflug, R.~R. Newton, S.~Peyton~Jones, and A.~Spiwack.
\newblock Linear haskell: Practical linearity in a higher-order polymorphic
  language.
\newblock {\em Proc. ACM Program. Lang.}, 2(POPL):5:1--5:29, 2017.

\bibitem{expoly}
L.~Cardelli and P.~Wegner.
\newblock On understanding types, data abstraction, and polymorphism.
\newblock {\em ACM Comput. Surv.}, 17(4):471--523, 1985.

\bibitem{rust}
The Rust~Project Developers.
\newblock {\em The Rust Programming Language}.
\newblock 2011.
\newblock \url{https://doc.rust-lang.org/book}.

\bibitem{rustinomicon}
The Rust~Project Developers.
\newblock {\em The Rustinomicon}.
\newblock 2011.
\newblock \url{https://doc.rust-lang.org/beta/nomicon/}.

\bibitem{ecmascript}
{ECMAScript} 2017 language specification (8th ed.), 2017.
\newblock \url{https://www.ecma-international.org/ecma-262/}.

\bibitem{l3-with-regions}
M.~Fluet, G.~Morrisett, and A.~Ahmed.
\newblock Linear regions are all you need.
\newblock In {\em Proc. European Symp. Programming}, pages 7--21, 2006.

\bibitem{compartments}
Andreas Gal.
\newblock Compartments, 2010.
\newblock \url{https://andreasgal.com/2010/10/13/compartments/}.

\bibitem{girard}
J.-Y. Girard.
\newblock Linear logic.
\newblock {\em Theoretical Computer Science}, 50(1):1--102, 1987.

\bibitem{rust-gc}
M.~Goregaokar and M.~Layzell.
\newblock Designing a {GC} in {Rust}, 2015.
\newblock
  \url{https://manishearth.github.io/blog/2015/09/01/designing-a-gc-in-rust/}.

\bibitem{jls}
J.~Gosling, B.~Joy, G.~Steele, G.~Bracha, A.~Buckley, and D.~Smith.
\newblock The {Java} language specification ({SE} 9 ed.), 2017.
\newblock \url{https://docs.oracle.com/javase/specs/jls/se9/html/}.

\bibitem{josephine}
Josephine, 2017.
\newblock \url{https://docs.rs/josephine/}.

\bibitem{cyclone}
T.~Jim, J.~G. Morrisett, D.~Grossman, M.~W. Hicks, J.~Cheney, and Y.~Wang.
\newblock Cyclone: A safe dialect of {C}.
\newblock In {\em Proc. {USENIX}}, pages 275--288, 2002.

\bibitem{K+R}
B.~W. Kernighan and D.~M. Ritchie.
\newblock {\em The C Programming Language}.
\newblock 2nd edition, 1988.

\bibitem{servo-gc}
J.~Matthews and K.~McAllister.
\newblock Javascript: {Servo}'s only garbage collector, 2014.
\newblock
  \url{https://research.mozilla.org/2014/08/26/javascript-servos-only-garbage-collector/}.

\bibitem{LISP}
J.~McCarthy.
\newblock Recursive functions of symbolic expressions and their computation by
  machine, part {I}.
\newblock {\em Commun. {ACM}}, 3(4):184--195, 1960.

\bibitem{linear-ocaml}
G.~Munch-Maccagnoni.
\newblock Resource polymorphism, 2018.
\newblock arXiv:1803.02796 [cs.PL].

\bibitem{spidermonkey}
Spidermonkey.
\newblock
  \url{https://developer.mozilla.org/en-US/docs/Mozilla/Projects/SpiderMonkey}.

\bibitem{servo}
Servo, the parallel browser engine project.
\newblock \url{https://servo.org/}.

\bibitem{regions}
M.~Tofte and J.-P. Talpin.
\newblock Region-based memory management.
\newblock {\em Inf. Comput.}, 132(2):109--176, 1997.

\bibitem{alms}
J.~A. Tov and R.~Pucella.
\newblock Practical affine types.
\newblock In {\em Proc. ACM Symp. Principles of Programming Languages}, pages
  447--458, 2011.

\bibitem{typescript}
Typescript.
\newblock \url{https://www.typescriptlang.org/}.

\bibitem{v8-embedding}
V8 embedder's guide.
\newblock \url{https://github.com/v8/v8/wiki/Embedder\%27s-Guide}.

\bibitem{walker}
D.~Walker.
\newblock Substructural type systems.
\newblock In B.~C. Pierce, editor, {\em Advanced Topics in Types and
  Programming Languages}, pages 3--43. MIT Press, 2002.

\end{thebibliography}

\end{document}